# Cascaded Second-order Surface Plasmon Solitons due to Intrinsic Metal Nonlinearity

Pavel Ginzburg[*], Alexey Krasavin, and Anatoly V. Zayats

Department of Physics, King's College London, Strand, London WC2R 2LS, United Kingdom

Abstract:

We theoretically show the existence of cascaded second-order surface plasmon solitons propagating at the interface between metal and linear dielectric. Nonlocal multipole nonlinearities originating from free conduction electron plasma of a metal lead to strong interaction between co-propagating surface plasmon polariton beams at fundamental and second harmonic frequencies. Finite element numerical modelling for effective two-dimensional medium explicitly demonstrates the solitons formation, confirming the theoretical results. The non-diffractive regime of propagation has been demonstrated at silica/silver interface for $5\lambda$ wide surface plasmon polariton beams with the loss-limited propagation distance of the order of 100 μm for the 750/1550 nm wavelengths pair. Plasmon-soliton formation in the phasematched conditions has been shown to be beneficial for nondiffractive surface plasmon polariton propagation.

[*]corresponding author:
pavel.ginzburg@kcl.ac.uk

1. <u>Introduction</u>

One of the most striking phenomena in nonlinear physics is the soliton formation. Solitons, wave packets, maintaining their shape (spatial or temporal) during their propagation, may emerge in various physical systems and were first discovered in narrow water channels [1]. However, considerable attention this general nonlinear phenomenon gained with the discovery of lasers, enabling the observation of soliton effects in optical domain [2,3]. Solitons can appear in quite general environment, where dispersion is compensated by nonlinearity, for example in solids (quantum-mechanical wavepackets) and many-particle biological systems [4]. Temporal solitons in optical fibres have particular applicative implementations as they enable dispersionless propagation of data bit streams along very long distances [3]. Spatial solitons may be used for pattern formation in semiconductor lasers and employed for data processing [5]. The second (lowest) order optical nonlinearity, under certain circumstances, can lead to the formation of temporal and spatial solitons [6].

Nano-optics is a very promising approach to achieve efficient nonlinear interactions, since it enables to manipulate electromagnetic modes in the near-field. One of the approaches for this control relies on subwavelength waveguiding and the associated field enhancement, based on surface plasmon polaritons (SPPs) waves, supported by materials with negative permittivity [7]. Nonlinear plasmonic phenomena, such as second-harmonic generation, cross- and self-modulation—all enhanced due to the plasmonic field, are sought after in active photonic components [8], sensing [9] and signal processing [10]. SPP propagation can be significantly affected by the nonlinearity of the adjacent medium. Nonlinear SPPs at the interface between metal and nonlinear Kerr dielectric in the presence of losses has been shown to result in self-focusing phenomenon with the formation of slowly decaying spatial solitons [11]. When SPPs are guided in layered structure of metal, nonlinear dielectric, and metal (MIM), hybrid vector spatial plasmon-solitons may emerge [12]. The most striking effect of plasmonics is that when the separation between the two metal claddings of the MIM structure is reduced in order to increase the transverse confinement into the deep subwavelength regime, the field envelope in the lateral dimension (bound only by the nonlinearity) is reduced as well [12]. This is opposite behaviour in comparison to nonlinear all-dielectric waveguides. A partial compensation of the propagation losses of surface plasmon soliton (SPS) in MIM structure was proposed at [13], where tapering at properly chosen angle leads to the additional field enhancement thus enabling longer soliton propagation. The combination of gain and loss media, adjusted to metal films can lead to stable spatial plasmon-solitons formations [14].

While many of the previous studies considered SPP self-focusing phenomena and soliton formation owing to adjacent nonlinear material, metals themselves are very nonlinear. The Kerr-like ponderomotive nonlinearity of noble metals at infrared spectrum was shown to be comparable with the highly nonlinear crystals [15]. Parametric optical processes in metals results from both the local interband based nonlinearity and the free electron plasma related nonlinearity, and, in most cases, from their combination. Metal interfaces are advantageous also for observation of second-order nonlinear processes requiring broken spatial symmetry. The required conditions are simply fulfilled by geometric violation of reflection symmetry at the metal-dielectric interface where SPPs reside. Therefore, there is an opportunity to use the second-order nonlinear effects in metal to control SPP propagation.

Second-order nonlinearity ($\chi^{(2)}$) can lead to spatial solitary wave formation via the effect of second-harmonic generation, as was theoretically predicted [6] and experimentally demonstrated almost two decades ago in KTP crystal [16] and planar $LiNbO_3$ waveguide [17]. The principle behind such spatial solitons is the collinear propagation of two beams at first and second harmonics frequencies. These beams exchange their energies via second-order polarizability, which provide the maximal phase lag at the region of maximal intensity, coining the term "cascaded $\chi^{(2)}$ solitons". This nonlinear phenomenon provide vast of opportunities for applications and fundamental studies, as was discussed in the comprehensive review on this topic [18].

Here we demonstrate the concept of cascaded $\chi^{(2)}$ surface plasmon solitons propagating at the interface between a linear dielectric and a metal with the nonlinearity described by hydrodynamic equations for conduction electron motion, showing that the beam propagation can be described by the effective nonlinear Schrödinger equation. These theoretical predictions are supported by effective two-dimensional numerical model, demonstrating the solitons formation in an explicit way. Furthermore, the phasematching between first and second harmonics on the soliton formation has been considered, extending the theoretical description and revealing its benefits (smaller solitons widths). Additionally, considering the regime of realistic propagation losses, we showed that the solitons can still be supported.

## 2. Metal Nonlinear Polarizabilities

Hydrodynamic equations provide satisfactory description of electrons' dynamics in the conduction band of noble metals, such silver and gold. Traditionally, material susceptibilities, linear as well as nonlinear, are derived with the help of averaged quantities: electron density ($n = n_0 + n_1 e^{-i\omega t} + n_2 e^{-2i\omega t} + ...$) and average velocity ($v = v_1 e^{-i\omega t} + v_2 e^{-2i\omega t} + ...$). The basic result for linear response is the well-known Drude model, which fits experimental data considerably well away from plasma frequency and interband transitions. Additional higher-order corrections and introduction of additional terms such as quantum pressure and viscosity of electron gas may lead to spatial dispersion contributions and temperature dependence of appropriated optical constants (see Ref. 19 and references therein). Careful inclusion of losses and interband transitions in the framework of hydrodynamic model provides more comprehensive but complex formulation [20]. However, the basic result of the Sommerfeld free-electron model extension for nonlinear polarization $\overrightarrow{P}^{(2)}(\omega,\omega)$ is based on the derivations of Bloembergen et. al. [21], which is core for other advanced models.

Expanding the electromagnetic fields in terms of fundamental and higher harmonics $\overrightarrow{E} = \overrightarrow{E_1}e^{-i\omega t} + \overrightarrow{E_2}e^{-2i\omega t} + ...,$ $\overrightarrow{H} = \overrightarrow{H_1}e^{-i\omega t} + \overrightarrow{H_2}e^{-2i\omega t} + ...$ and substituting them into hydrodynamic equations, the basic second-harmonic polarization can be derived [21]:

$$\overrightarrow{P}^{(2)}{}_{NL}(\omega,\omega) = \frac{e\varepsilon_b \omega_p^2}{4m\omega^4}\left(\overrightarrow{E_1}\cdot\vec{\nabla}\right)\overrightarrow{E_1} + \frac{e\varepsilon_b}{2m\omega^2}\left(\nabla\cdot\overrightarrow{E_1}\right)\overrightarrow{E_1}, \tag{1}$$

where $\omega_p^2 = \dfrac{n_0 e^2}{m\varepsilon_b}$ is the electron plasma frequency, $\omega$ is the field angular frequency, $n_0$ is the unperturbed electron concentration, $m$ is the electron effective mass, and $\varepsilon_b$ is the background permittivity. This polarization term describes the frequency doubling. In the similar fashion, we derive the polarization, describing the down-conversion process:

$$\overrightarrow{P}^{(2)}{}_{NL}(2\omega,-\omega) = -\frac{e\varepsilon_b}{\omega^2 m}\left(\nabla\cdot\overrightarrow{E_2}\right)\overrightarrow{E_1}^* + \frac{e\varepsilon_b}{2\omega^2 m}\left(\nabla\cdot\overrightarrow{E_1}^*\right)\overrightarrow{E_2}. \tag{2}$$

Eqs. 1 and 2 reveal the possibility of the energy exchange between two propagating beams at the frequency of ω and 2ω through the second-order nonlinearity in metals.

### 3. Nonlinear Equations for General Cascaded SPS

In the linear regime, the SPP modes on the metal-dielectric interface are described by

$$
\begin{aligned}
&E_i(z,t) = A_i(y,z)\left[\hat{x}e_{ix} + i\hat{z}e_{iz}\right]e^{i\beta_i z - i\omega_i t}, \\
&\beta_i = \frac{\omega_i}{c}\sqrt{\frac{\varepsilon_m(\omega_i)\varepsilon_d}{\varepsilon_m(\omega_i)+\varepsilon_d}}, \\
&k_{id,m}^2 = \beta_i^2 - \frac{\omega_i^2}{c^2}\varepsilon_{d,m}, \\
&e_{ix} = \pm\frac{\beta_i}{k_i}e^{\mp k_i x}, e_{iz} = e^{\mp k_i x},
\end{aligned}
\tag{3}
$$

where *i=1,2* corresponds to the waves at fundamental and second harmonic frequencies, $A_i(y,z)$ is the spatial dependent mode amplitude, $\beta_i$ is the linear propagation constant, and $\varepsilon_{d,m}$ is the linear permittivity of dielectric and metal. The geometrical configuration is depicted on Fig. 1 (a,b). Taking into account that the SPP modes are TM-polarized, the nonlinear wave equation can be decomposed via x- and z- components and, subtracting the identities for linear propagation regime, we obtain

$$\partial_{yy}A_i e_{ix}e^{-i\beta_i z+i\omega_i t} - i\partial_z A_i e_{iz}e^{-i\beta_i z+i\omega_i t} - 2i\beta_i\partial_z A_i e_{ix}e^{-i\beta_i z+i\omega_i t} - \mu_0\partial_{tt}P_{NLx} = 0,$$
$$\partial_{yy}A_i e_{iz}e^{-i\beta_i z+i\omega_i t} + i\partial_z A_i e_{ix}e^{-i\beta_i z+i\omega_i t} + i\mu_0\partial_{tt}P_{NLz} = 0,$$

(4)

where $\partial_\alpha$ denotes the derivative with respect to subscript index $\alpha$ and $\mu_0$ is the vacuum permeability (nonmagnetic materials were assumed).

Multiplying the set of Eq. 4 by $e_{ix}$ and $e_{iz}$, respectively, and summing up the results taking into account the identity for waveguide modes ($e_{ix}\partial_x e_{iz} = e_{iz}\partial_x e_{ix}$), we arrive to

$$\partial_{yy}A_i\left(e_{ix}^{\,2}+e_{iz}^{\,2}\right)e^{-i\beta_i z+i\omega_i t} - 2i\beta_i\partial_z A_i e_{ix}^{\,2}e^{-i\beta_i z+i\omega_i t} - \mu_0\partial_{tt}P_{NLx}e_{ix} + i\mu_0\partial_{tt}P_{NLz}e_{iz} = 0.$$

(5)

This is the general expression for nonlinear TM modes, similar to the one derived in [11]. Integration over the transverse direction of the waveguide results in

$$\partial_{yy}A_i\left[\int_{-\infty}^{\infty}\left(e_{ix}^{\,2}+e_{iz}^{\,2}\right)dx\right]e^{-i\beta_i z+i\omega_i t} - 2i\beta_i\partial_z A_i\left[\int_{-\infty}^{\infty}e_{ix}^{\,2}dx\right]e^{-i\beta_i z+i\omega_i t} - \mu_0\partial_{tt}\int_{-\infty}^{\infty}\left(P_{NLx}e_{ix}-iP_{NLz}e_{iz}\right)dx = 0$$

(6)

This set of equations (*i=1,2*) describes the self-consistent process of energy exchange between two propagating beams, coupled via any general nonlinear polarizability term.

4. Cascaded SPPs, originated from hydrodynamic nonlinearity

Having derived Eq. 6, the next task is to incorporate the actual nonlinear polarizabilities given by of Eqs. 1 and 2. First, we observe that for TM modes $\left(\vec{E}\cdot\vec{\nabla}\right)\vec{E} = \left(\nabla\cdot\vec{E}\right)\vec{E}$, simplifying the subsequent derivations. It can be seen that $\nabla\cdot\vec{E}$ is only non-vanishing at the boundary and, in fact, provides the measure of the polarization charge density at the interface between the metal and dielectric. The integration of Eq. 6 will eliminate the delta-function, corresponding to the surface charge density and will retain the field values in the metal, next to the boundary. Typically, far from surface plasmon frequency, the z-component (longitudinal) of the SPP field in metal dominates the transverse in $\sim |\varepsilon_m|$ times (2 orders of magnitude in the infrared spectral range) and, hence, may be kept as the source of the leading term in the nonlinear polarization with the other components neglected (they can be taken into account with the mere result of algebraic complications). Defining the nonlinear coefficients $\alpha = \dfrac{e\varepsilon_b \omega_p^2}{4m\omega^4} + \dfrac{e\varepsilon_b}{2m\omega^2}$ and $\beta = \dfrac{e\varepsilon_b}{2m\omega^2}$, the resulting equations for the fundamental and second- harmonic SPPs can be written as

$$\partial_{yy} A_1 \left[ \int_{-\infty}^{\infty} \left( e_{1x}^2 + e_{1z}^2 \right) dx \right] e^{-i\beta_1 z + i\omega t} - 2i\beta_1 \partial_z A_1 \left[ \int_{-\infty}^{\infty} e_{1x}^2 dx \right] e^{-i\beta_1 z + i\omega t} + i\mu_0 \omega^2 \beta e_{2z} |e_{1z}|^2 A_1^* A_2 e^{i(\beta_1 - \beta_2)z + i\omega t} = 0,$$
$$\partial_{yy} A_2 \left[ \int_{-\infty}^{\infty} \left( e_{2x}^2 + e_{2z}^2 \right) dx \right] e^{-i\beta_2 z + i2\omega t} - 2i\beta_2 \partial_z A_2 \left[ \int_{-\infty}^{\infty} e_{2x}^2 dx \right] e^{-i\beta_2 z + i2\omega t} + i4\mu_0 \omega^2 e_{1z} e_{1z} e_{2z} \alpha A_1^2 e^{-2i\beta_1 z + i2\omega t} = 0. \tag{7}$$

Introducing abbreviations $I_i = \int_{-\infty}^{\infty} \left( e_{ix}^2 + e_{iz}^2 \right) dx$, $I_{Xi} = \int_{-\infty}^{\infty} e_{ix}^2 dx$, $N_1 = i\mu_0 \omega^2 \beta e_{2z} |e_{1z}|^2$,

$N_2 = i4\mu_0 \omega^2 e_{1z}^2 e_{2z}$, and $\Delta = 2\beta_1 - \beta_2$ we rearrange the set of Eq. 7 as

$$\partial_{yy} A_1 I_1 - 2i\beta_1 \partial_z A_1 I_{X1} + N_1 A_1^* A_2 e^{i\Delta z} = 0,$$
$$\partial_{yy} A_2 I_2 - 2i\beta_2 \partial_z A_2 I_{X2} + N_2 A_1^2 e^{-i\Delta z} = 0. \tag{8}$$

Assuming also large momentums mismatch between fundamental and second-harmonic SPPs (approximation of [22]) which is usually justified due to the SPPs dispersion, we can show that the second equation of the set of Eq. 8 will result in

$$A_2 = \frac{N_2 e^{-i\Delta z}}{2\beta_2 \Delta I_{X2}} A_1^2. \tag{9}$$

Substituting this amplitude back into the first of Eq. 8, we obtain:

$$\partial_{yy} A_1 I_1 - 2i\beta_1 \partial_z A_1 I_{X1} + \frac{N_2 N_1}{2\beta_2 \Delta I_{X2}} |A_1|^2 A_1 = 0, \tag{10}$$

which is the final result of the derivations. Eq. 10 is the nonlinear Schrödinger equation describing the nonlinear propagation of the first-harmonic (fundamental) SPP mode on the surface of the metal exhibiting second-order nonlinearity with a possible solution corresponding to solitary wave. It is the most general tool for soliton description, and its coefficients indicate whether soliton can emerge. Even though the derivations rely on large momentum mismatch, the final result for the effective nonlinear coefficient in Eq. 10 singularly grows close to the phasematching condition, indicating its possible advantage.

5. <u>Simulation results</u>

To further investigate the nonlinear interaction between the SPP signals finite element numerical simulations were employed. SPP modes are strongly confined to the metal-dielectric interface where the nonlinear interactions take place. For the sake of numerical simplicity, we have investigated general two-dimensional model of cascaded $\chi^{(2)}$ solitons, for which the signal dependence on the *x*-coordinate was omitted, qualitatively not affecting the results since *x* is a dummy variable in Eq. 10. In the same time, the mismatch between the effective refractive indexes for the fundamental and the double-frequency signals reflecting the dispersion of the plasmonic waves and determining the essential phase relations between the waves and was taken into account. Furthermore, an additional imaginary term was added to the medium effective refractive indexes, taking into account the attenuation of the SPP waves. In the first studied case of silica/silver interface ( $\varepsilon_1^{Ag} = -120 + 3i$ , $\varepsilon_2^{Ag} = -27 + 0.32i$ at $\lambda_1 = 1500$ nm and $\lambda_2 = 750$ nm , respectively [23]) the obtained refractive indexes for the simulated effective medium for the first and the second harmonics are $n_1^{eff} = 1.457 + 3.25 \cdot 10^{-4} i$ and $n_2^{eff} = 1.514 + 7.5 \cdot 10^{-4} i$ , corresponding the mismatch between the wavevectors of approximately 3.8% and the SPP propagation length 370 μm and 80 μm, respectively. At the source boundary (*z*=0), SPP beams at fundamental frequency, corresponding to the free-space wavelength $\lambda_1 = 1500$ nm , and second-harmonic one ( $\lambda_2 = 750$ nm ) were set to propagate collinearly along *z*-axis with the transverse profile described by the Gaussian distributions $E_{1,2}(y,0) = A_{1,2}(y,0) = \exp(-y^2/w_{1,2}^2)$ with half-widths of $w_{1,2} = 2.5\lambda_{1,2}$. Their spatial evolution was then studied.

The intensity distributions $|E_{1,2x}|^2$ obtained in the linear (uncoupled) regime show typical diffraction-governed propagation for both fundamental and second-harmonic frequency SPP beams (Fig. 2). The intensity distributions in the nonlinear propagation regime were simulated with nonlinear polarization terms $P_{1,NL} = \chi^{(2)}(E_1)^* E_2$ and $P_{2,NL} = 1/2 \cdot \chi^{(2)}(E_1)^2$ assuming the source field amplitudes for both harmonics identical $A_1 = A_2$. It should be noted as the amplitudes of the fundamental and second-order SPP waves are connected in the solitonic regime via Eq. (9), the choice of the relative amplitudes of the two waves are not important as they adjust themselves to reach the required ratio. The nonlinear interaction was gradually increased to $\chi^{(2)}E_1 = 0.02$ resulting in the intensity maps in Fig. 2 (b,f) where deviations from the linear propagation regime is seen. Further increase of the field to $\chi^{(2)}E_1 = 0.05$ leads to even more pronounced deviations from diffractive regime (Fig. 2 (c,g)). The observed intensity fringes are defined by the mismatch between the effective refractive indexes at the two frequencies. In the same time, the energy exchange between the beams can be observed there: maximum intensity of one beam corresponds to the minimum intensity of the other (compare cross sections $A$ and $B$ in maps (c) and (f)). Same can be seen in the actual intensity plots along these cross sections (graphs (d) and (h)). Furthermore, the effect of narrowing of the beams, driven by the nonlinearity can be seen in the decrease in the average beam width $(w_{1,B} + w_{1,C})/2/w_{1,A} = 0.94$ for the fundamental frequency and $(w_{2,B} + w_{2,C})/2/w_{2,A} = 0.8$ for the double frequency, where $w_B$ and $w_C$ are the interchanging maximum/minimum of SPP beam widths in the solitonic regime, and $w_A$ is the beam width at the linear (non-interacting) propagation regime at the same average distance. It can be noted, that largest modulation of the field profiles is in the centres of the beams, where intensities are the highest and nonlinear coupling is the strongest. Also, the initial choice of the beam widths was optimized to fit the solitonic regime, while other width ratios (e.g. equal beam widths in microns) were checked to lead for self-focusing or, on the other hand, diffraction.

One of the assumptions, made prior to derivation of Eq. 9 is the large wavevector mismatch between the first and second harmonic SPPs. At the same time, the nonlinear coefficient $N_1 N_2 / (2 \beta_2 \Delta I_{X_2})$ in this equation has the mismatch term in the denominator and hence larger for smaller mismatches. In these circumstances, a particular interest presents the scenario when the mismatch is zero so that phase-matching conditions realised. This can be obtained by introducing at the SPP-supporting interface dielectric having anomalous dispersion [24] to compensate SPP dispersion. Adjusting glass composition to 5TiO$_2$-55SiO$_2$-40Na$_2$O it is possible to achieve a match between the effective refractive indexes $n_1^{eff} = n_2^{eff} = 1.583$. The evolution of the plasmonic beams for linear and nonlinear cases at this interface was further studied. The resulting intensity maps organized in the same layout as for the silica/silver material pair are presented in Fig. 3. On these maps one can easily see the rigorous transformation of the plasmonic modes into highly-localized non-diffracting solitons. Moreover, the nonlinearity actually spatially compress the plasmonic fields into more narrow transverse profile, which can be seen in the self-focusing effect occurring near the source boundary (Fig. 3 (c) and (f)).

6. <u>Summary</u>

We have derived the nonlinear Schrödinger equation describing the SPP soliton formation in cascaded second-order nonlinearity regime. The nonlinearity originates from the presence of the metal interface and derived in the hydrodynamic description of free-electron plasma. Finite element simulations performed with rigorously introduced second-order coupling between the fundamental and the second harmonic beams confirm the formation of cascaded plasmon-solitons, underlining benefits of phasematching for non-diffractive SPP propagation over distances of hundreds of microns.

Acknowledgments

This work has been supported in part by EPSRC (UK). P. G. acknowledges the Royal Society for Newton International Fellowship. Authors thank Prof. Dmitry Skryabin for discussions.

Figure Captions:

Figure 1. (a) Schematic of a suggested experimental setup: light at fundamental and second-harmonic frequencies is coupled to the SPPs co-propagating on the interface between metal and linear dielectric. (b) Surface plasmon polariton cascaded $\chi^{(2)}$ soliton geometry: co-propagating SPP beams at fundamental and second-harmonic frequency can be coupled via the metal nonlinearity. The field distributions of the SPP beams are shown.

Figure 2. Linear propagation of (a) fundamental and (e) second-harmonic SPP beams in effective two-dimensional medium (Au/silica interface) with the effective indexes $n_1^{eff} = 1.457$ and $n_2^{eff} = 1.514$ at 1500 nm and 750 nm wavelengths, respectively. Nonlinear propagation and selffocusing of (b,c) fundamental- and (f,g) second-harmonic SPP beams for different light intensities corresponding to (b,f) $\chi^{(2)}E_1 = 0.02$ and (c,g) $\chi^{(2)}E_1 = 0.05$ nonlinearity. The initial amplitudes of the both beams are equal ($E_1 = E_2$) and the beam half-widths are $w_{1,2} = 2.5\lambda_{1,2}$. The graphs (d) and (g) show the intensity plots along cross sections indicated in (c) and (g), respectively.

Figure 3. Linear propagation of (a) fundamental- and (d) second-harmonic SPP beams in two-dimensional medium with the equal effective indexes $n_1^{eff} = n_2^{eff} = 1.583$. Nonlinear propagation and self-focusing of (b,c) fundamental- and (e,f) second-harmonics SPP beams for different light intensities corresponding to (b,e) $\chi^{(2)}E_1 = 0.02$ and (c,f) $\chi^{(2)}E_1 = 0.05$. The amplitudes of the both beams are equal ($E_1 = E_2$) and the beam half-widths are $w_{1,2} = 2.5\lambda_{1,2}$.

Fig. 1

(a)

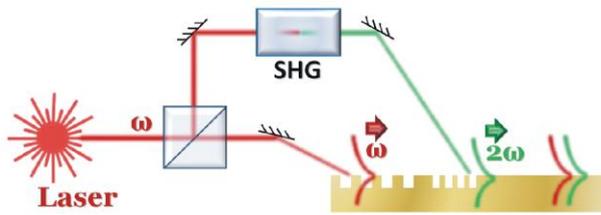

(b)

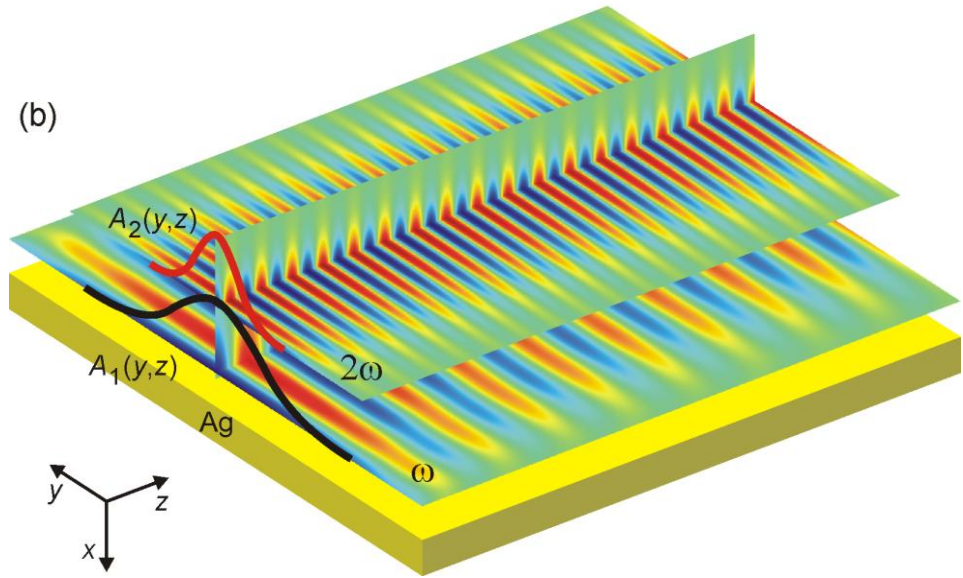

Fig. 2

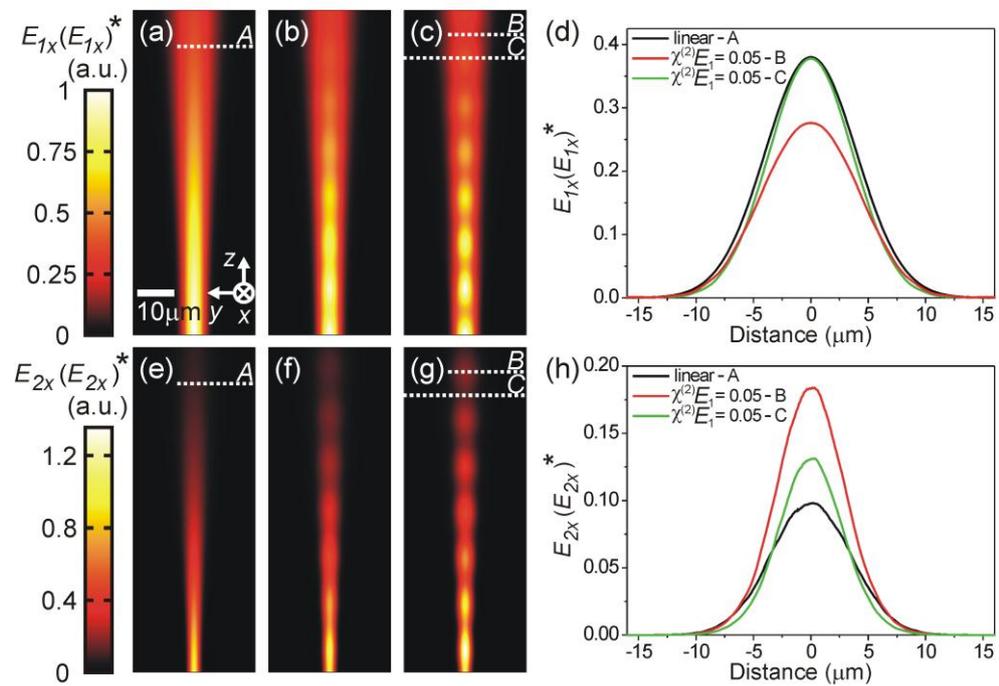

Fig. 3

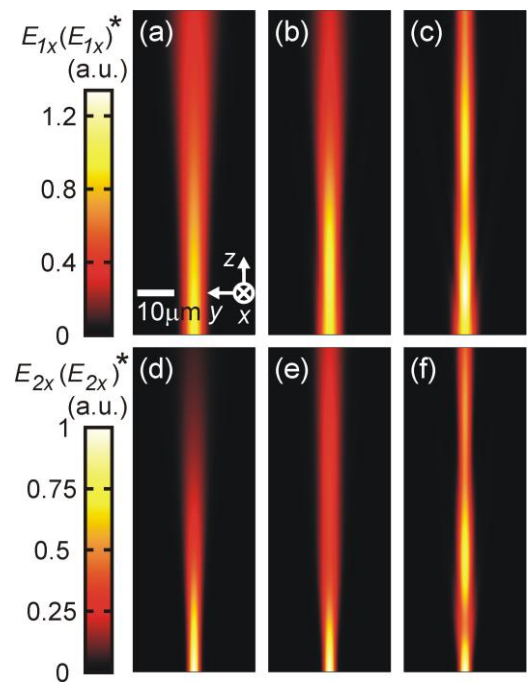